\documentclass[twocolumn,superscriptaddress,amsmath,amssymb,aps]{revtex4-1}

\usepackage{graphicx}
\usepackage{dcolumn}
\usepackage{bm}
\usepackage{multirow}
\usepackage{setspace}
\usepackage{hyperref}
\usepackage{cleveref}
\usepackage{lipsum}
\usepackage{booktabs}
\usepackage{amsmath}
\usepackage{mathrsfs}
\usepackage{subfigure}
\usepackage{scrextend}
\usepackage{physics}
\usepackage{braket}
\usepackage{scrextend}

\begin{document}

\title{High-resolution conversion electron spectroscopy of the $^{125}$I electron-capture decay}

\author{B.P.E.~Tee}
\affiliation{Department of Nuclear Physics, Research School of Physics and Engineering, The Australian National University, Canberra, ACT, Australia}
\author{A.E.~Stuchbery}
\affiliation{Department of Nuclear Physics, Research School of Physics and Engineering, The Australian National University, Canberra, ACT, Australia}
\author{M.~Vos}
\affiliation{Department of Electronic Materials Engineering, Research School of Physics and Engineering, The Australian National University, Canberra, ACT Australia}
\author{J.T.H.~Dowie}
\affiliation{Department of Nuclear Physics, Research School of Physics and Engineering, The Australian National University, Canberra, ACT, Australia}
\author{B.Q.~Lee}
\affiliation{Department of Nuclear Physics, Research School of Physics and Engineering, The Australian National University, Canberra, ACT, Australia}
\author{M.~Alotiby}
\affiliation{Department of Electronic Materials Engineering, Research School of Physics and Engineering, The Australian National University, Canberra, ACT Australia}
\affiliation{King Abdulaziz City for Science and Technology, Riyadh, Saudi Arabia}
\author{I.~Greguric}
\affiliation{Australian Nuclear Science and Technology Organisation, Lucas Heights, NSW,
	Australia}
\author{T.~Kib\'{e}di }
\email[]{Tibor.Kibedi@anu.edu.au}
\affiliation{Department of Nuclear Physics, Research School of Physics and Engineering, The Australian National University, Canberra, ACT, Australia}

\date{\today}

\begin{abstract}
	\noindent The conversion electrons from the decay of the 35.5-keV excited state of $^{\rm 125}$Te following the electron capture decay of $^{\rm 125}$I have been investigated at high resolution using an electrostatic spectrometer. The penetration parameter $\lambda = -1.2(6)$ and mixing ratio $\abs{\delta(E2/M1)}=0.015(2)$ were deduced by fitting to literature values and present data. The shake probability of the conversion electrons is estimated to be 0.5, more than two times larger than the predicted value of 0.2.	
 \vspace*{12pt} \\
   (Accepted by Physical Review C, 29 August 2019)

\end{abstract}

\maketitle

\section{Introduction}
 The probability of the emission of a conversion electron is most often evaluated from the probability of $\gamma$-ray emission and the internal conversion coefficient (ICC), $\alpha$. This assumes that all nuclear structure effects are contained in the $\gamma$-ray emission probability and $\alpha$ only depends on atomic properties. In this case, the interaction between the conversion electron and the nucleus only takes place outside the nucleus \cite{church1960nuclear}. This picture is valid for most transitions; however, the atomic electron involved in the conversion process may penetrate into the nucleus and interact with the transition charges and currents in the interior of the nucleus. The corresponding \emph{``dynamic penetration"} matrix elements are dependent on nuclear structure and not necessarily proportional to the  $\gamma$-ray matrix elements (as in the case of the point-like nucleus assumption). These finite nucleus effects may result in anomalies in the measured ICCs, i.e deviations from theory which assumes point-like nuclei. 
 The resulting conversion coefficient of (sub-)shell $i$, $\alpha_i$, can be expressed in terms of the ``unperturbed"
 conversion coefficients $\alpha_i(ML)$, for pure magnetic multipoles \cite{pauli1967finite},
 \begin{equation}
   \alpha_i = \alpha_i(ML)(1+b_1(i) \lambda + b_2(i) \lambda^2),
  \label{eqn:ICC_ML}
\end{equation}
and $\alpha_i(EL)$, for pure electric multipoles \cite{pauli1967finite},
\begin{equation}
\begin{split}
   \alpha_i = \alpha_i(EL)(1+a_1 (i)\eta + a_2 (i)\eta^2  \\
    +a_3(i)\eta\zeta + a_4 (i)\zeta + a_5 (i)\zeta^2 ) \, ,
  \label{eqn:ICC_EL}
\end{split}
\end{equation}
 where $b_1(i)$, $b_2(i)$, $a_1(i)$, $a_2(i)$, $a_3(i)$, $a_4(i)$ and $a_5(i)$ are theoretical
 penetration coefficients of sub-shell $i$, which depend only on the electron wavefunctions,
 and $\lambda$, $\eta$ and $\zeta$ are the penetration
 parameters containing nuclear structure information independent of the atomic shell.
For magnetic dipole transitions, the dimensionless penetration parameter is defined to be \cite{church1960nuclear,berghe1970penetration}
\begin{equation}
\begin{split}
\lambda &\equiv \frac{\bra{f|}M_{e}\ket{|i}}{\bra{f|}U_\gamma(M1)\ket{|i}} \\
&\approx \frac{1}{1+\sigma_{C}} \frac{\bra{f|}M_{e1}\ket{|i} + \sigma_{C} \bra{f|}M_{e2}\ket{|i}}{\bra{f|}U_\gamma(M1)\ket{|i}},
\label{eq:lambda_defined}
\end{split}
\end{equation}
\noindent where $\bra{f|}M_{e1}\ket{|i}$ and $\bra{f|}M_{e2}\ket{|i}$ are the penetration matrix elements, $\bra{f|}U_\gamma(M1)\ket{|i}$ is the usual $M1$ $\gamma$-radiation matrix element and $\sigma_{C}$ is an expansion coefficient that is generally small (with magnitude between 0.1 to 0.2). The $\sigma_{C}$ values are given in \cite{church1966tables}.

In a particle-core coupling model, the operators are formulated as follows \cite{berghe1970penetration}:
\begin{equation}
\begin{split}
M_{e1} &= A (g_l\bm{l} + \frac{5}{3}g_s^{(a)}\bm{s} + g_R \bm{R} \\
       &+ \frac{2}{3}\sqrt{2\pi}g_s^{(b)}[Y_2 \otimes \bm{s}]^{(1)})(\frac{r}{R_0})^2 \\
M_{e2} &= A (g_l\bm{l} + \frac{7}{3}g_s^{(a)}\bm{s} + g_R \bm{R} \\
       &+ \frac{4}{3}\sqrt{2\pi}g_s^{(b)}[Y_2 \otimes \bm{s}]^{(1)})(\frac{r}{R_0})^4 \\
U_{\gamma}(M1) &= A (g_l\bm{l} + g_s^{(M1)}\bm{s} + g_R\bm{R}),
\label{eq:PE_GAMMA_MATRIX}
\end{split}
\end{equation}
\noindent where $A = (\frac{3}{4\pi})^{1/2}\frac{e\hbar}{2Mc}$, $\bm l$ and $\bm s$ are the orbital and spin angular momentum operators for the odd nucleon and $\bm R$ is the angular momentum operator of the core. For $l$-forbidden transitions the tensor term $[Y_2 \otimes \bm{s}]^{(1)}$ dominates. In this term $(r/R_0)$ represents the reduced radial co-ordinate, where $R_0$ is the nuclear radius taken as 1.2A$^{\rm 1/3}$ fm. The gyromagnetic ratios $g_l$, $g_s$, $g_R\approx$ Z/A are associated with the orbital, spin and collective motions, respectively. Note the effective $g_s$ factors: $g_s^{(a)}$, $g_s^{(b)}$ and $g_s^{(M1)}$ are expected to be of the order of the free nucleon $g_s$ value but could be different in general, and $M_{e1}$ and $M_{e2}$ are distinctly different from $U_\gamma(M1)$ by the appearance of the tensor term $[Y_2 \otimes \bm{s}]^{(1)}$ and the overall radial integral factor. In this paper for brevity we shall write the matrix elements $\bra{f|}Q\ket{|i}$ as $Q$, where $Q$ = $U_\gamma$, $M_{e1}$ or $M_{e2}$.
 \begin{figure}
	\centering
	\includegraphics[width=0.7\columnwidth]{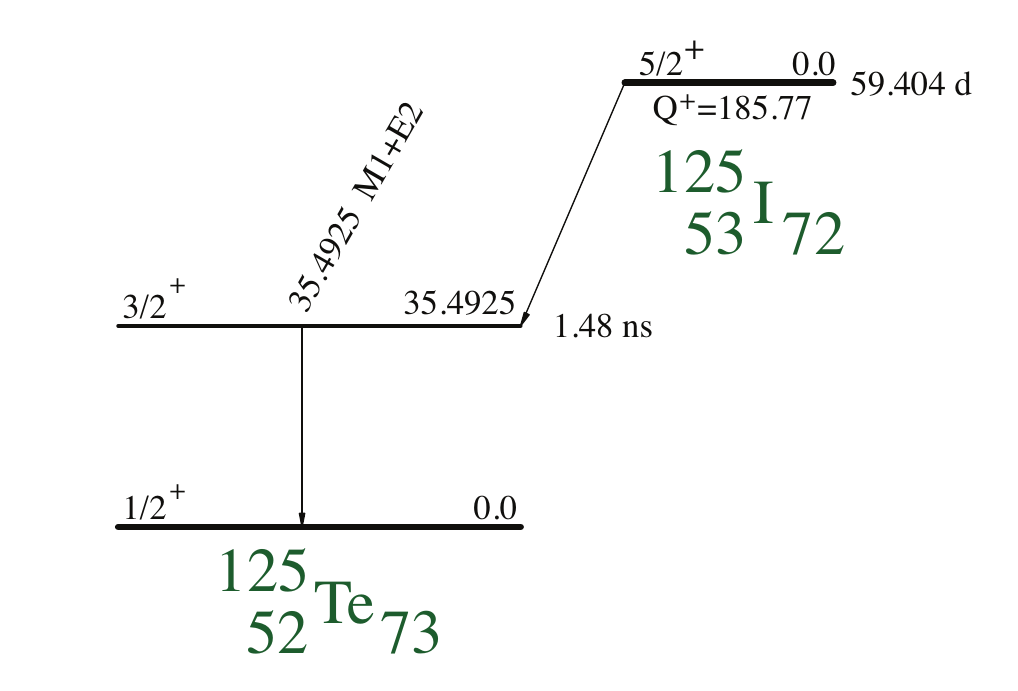}
	\vspace*{-9pt}
	\caption{The decay scheme of $^{125}$I.}
	\label{fig:125I_EC_scheme}
\end{figure}
 The decay of the 35.5-keV excited state of $^{\rm 125}$Te following the electron capture (EC) decay of $^{\rm 125}$I is one of the few cases where the nuclear structure effect could affect the ICCs.

 The decay scheme of $^{125}$I into the $^{125}$Te is shown in Fig.~\ref{fig:125I_EC_scheme}. The $5/2^{+}$ ground state of $^{125}$I decays with an allowed EC transition to the 35.5-keV $3/2^{+}$ excited state in the $^{125}$Te daughter nucleus. The direct EC decay to the $1/2^{+}$ ground state of $^{125}$Te with a second forbidden $\Delta J=2$ transition is highly retarded; its probability is less than 1\% of the total decay intensity \cite{1966Sm05}. To EC decay to the second excited state in $^{125}$Te at 144.775 keV and $J^{\pi}=11/2^{-}$ would require a third forbidden transition with $\Delta J =3$; it is very unlikely and not observed.

 The 35.5-keV $3/2^{+}$ excited state decays electromagnetically to the $1/2^+$ ground state of ${^{125}}$Te through a $\gamma$-transition or the emission of a conversion electron. The selection rule for $\gamma$-transitions and previous studies indicate that the 35.5-keV transition is a mixed $M1$+$E2$ transition, dominantly of $M1$ multipolarity \cite{KATAKURA2011495}. The conversion coefficient for a mixed $M1$+$E2$ transition is related to $\alpha(ML)$ and $\alpha(EL)$ by
 \begin{equation}
 \alpha(M1/E2) = \frac{\alpha(M1)+\delta^2 \alpha(E2)}{1 + \delta^2},
 \label{eq:Mixing}
 \end{equation}
 \noindent where the square of the multipole mixing ratio $\delta^2$ is the ratio of the $E2$ and $M1$ $\gamma$-transition rates. The sign of $\delta$ follows the convention by Krane and Steffen \cite{krane1970determination} and $\delta$ is given by:
 \begin{equation}
 \delta = 0.835 E_\gamma\frac{\bra{f|}E2\ket{|i}}{\bra{f|}M1\ket{|i}},
 \label{eq:mixing_ratio}
 \end{equation}
 \noindent where $E_\gamma$ is the transition energy in MeV, and $\bra{f|}E2\ket{|i}$ and $\bra{f|}M1\ket{|i}$ are the reduced matrix elements of the $E2$ and $M1$ operators. The $M1$ $\gamma$-transition in $^{125}$Te is $l$-forbidden because the change in orbital angular momentum of the two states involved is $\Delta l$=2 ($\nu_{d_{3/2}} \rightarrow \nu_{s_{1/2}}$). Thus the $\gamma$-ray matrix element $U_\gamma(M1)$ is expected to be small while the penetration matrix elements $M_{e1}$ and $M_{e2}$ are allowed due to the tensor terms. Hence the small $U_\gamma(M1)$ with finite $M_{e1}$ and $M_{e2}$ may result in non-negligible $\lambda$ values, causing anomalies in the observed ICCs. This scenario was first suggested by Church and Weneser \cite{church1960nuclear} in 1960.

 The study of anomalous ICCs resulting from the penetration effect provides an opportunity to test nuclear structure models by comparing the calculated penetration parameter $\lambda$ with experiment. In the theoretical calculations of the penetration matrix elements, $M_{e1}$ and $M_{e2}$ are approximately proportional to the spin gyromagnetic ratio $g_s^{(b)}$ for the $l$-forbidden $M1$ $\gamma$ transitions. The measurements of $\lambda$ could therefore be used to deduce the $g_s^{(b)}$ factor, which is expected to differ from the $g_s$ factor for free nucleons (where $g_s^{\rm free}(p)=5.58$ for proton, and $g_s^{\rm free}(n)=-3.83$ for neutron) and may also differ from $g_s^{(M1)}$ which affects $M1$ transition rates and magnetic moments. Hence the renormalization of the $g_s^{(b)}$ factors can be observed in the anomalous ICC measurements and, in turn, this gives information on the spin-force constants \cite{listengarten1976renormalization,listengarten1978penetration}.

 $^{125}$I is a commonly-used medical isotope. Measuring low-energy electron spectra at high precision has been part of our program to improve the knowledge of atomic radiations, including Auger electrons, for medical isotopes \cite{alotiby2018measurement,ALOTIBY2019}. The measurements determine an accurate absolute Auger electron yield from a radioisotope by the simultaneous measurement of conversion and Auger electrons. Precise knowledge on the conversion electrons is thus required. Here we report on a measurement of the intensity ratios of conversion electrons from the $l$-forbidden 35.5-keV transition in $^{125}$Te.

\section{Experiments}
 Sub-monolayer films of radioactive $^{125}$I atoms were deposited on a Au(111) surface following the procedure described by Pronschinske \textit{et al.} \cite{pronschinske2015enhancement}. The $^{125}$I activity was obtained commercially from PerkinElmer (part number: NEZ033A002MC) and the sources were prepared at the Australia's Nuclear Science and Technology Organisation (ANSTO). The Au(111) surface was obtained by flame annealing the Au samples just before the $^{125}$I deposition. A droplet containing Na$^{125}$I in a 100 $\rm \mu$L 0.02 M NaOH solution (pH $\approx$ 10) was put on this surface, and left to react. An approximately 4 mm diameter source with an activity of 5 MBq was obtained in this way.

 \begin{figure}
	\centering
	\includegraphics[width=1\columnwidth]{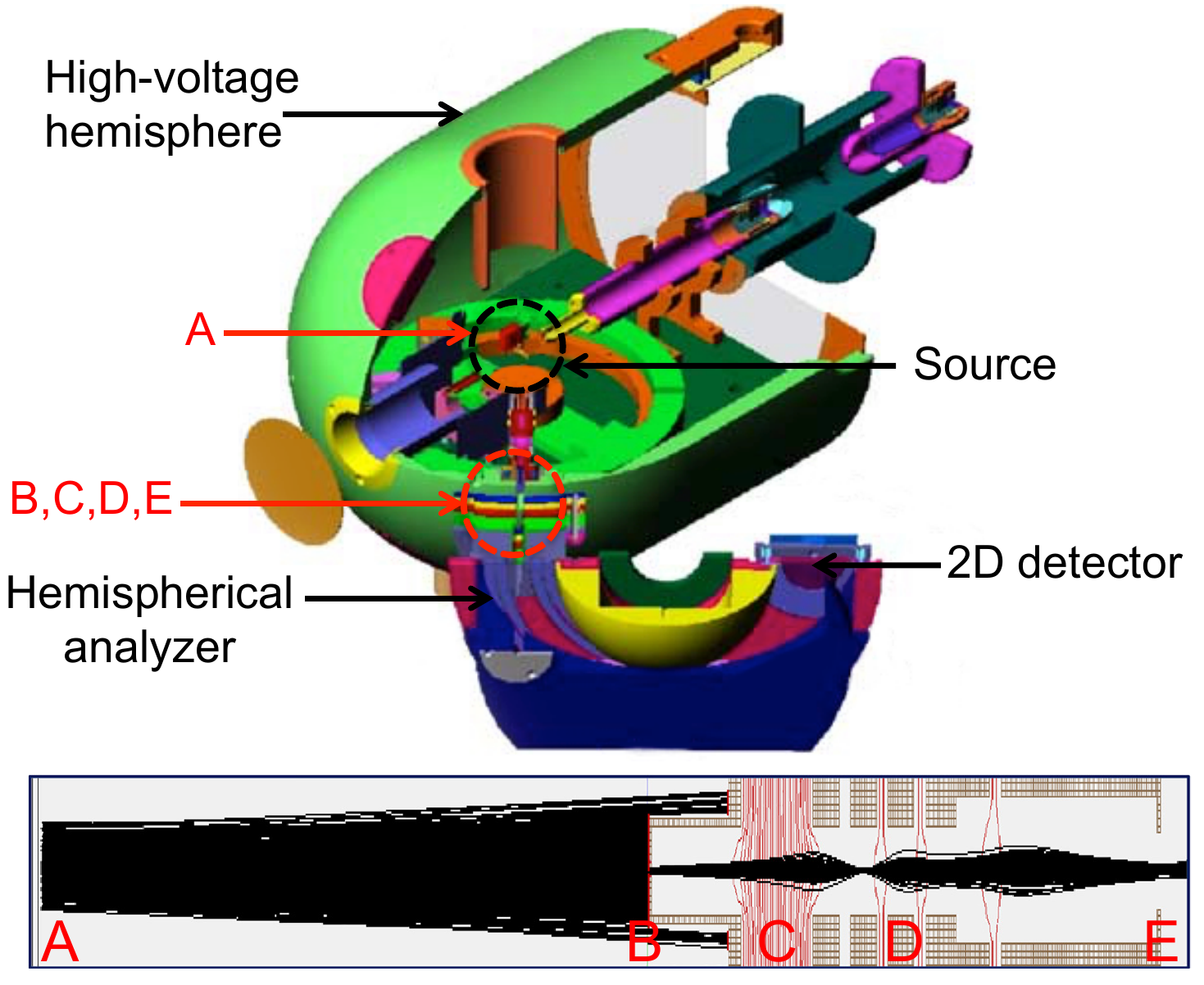}
	\caption{Top panel: Layout of the electrostatic spectrometer \cite{vos2000high,went2005electron}. The details are described in the main text.  Bottom panel: SIMION simulations for the electrostatic spectrometer. Electrons emitted from a 4 mm diameter source (A) are restricted from entering the lens stack by a 0.5 mm wide slit (B). After the main deceleration stage (C) the electrons are focused by a set of electrostatic lenses (D) on the exit plane (E) that coincides with the entrance of the hemispherical analyzer. The vertical scale is expanded by a factor of 4 for clarity, and the red lines are equipotential planes plotted at 1000 V intervals.}
	\label{fig:EMS_SIMION}
\end{figure}

 The conversion electron measurements were performed using an electrostatic spectrometer
 that is capable of measuring electrons with energies from 2 keV up to 40 keV
 \cite{vos2000high,went2005electron}. A layout of the electrostatic spectrometer is presented in Fig.~\ref{fig:EMS_SIMION} along with a simulation of the electron transport through the spectrometer.
 The sample was held at the center of a positive high voltage hemisphere in an ultra high vacuum ($\approx 10^{-10}$ mbar) chamber.
 At the exit of the hemisphere, the emitted electrons are collimated by passing through
 a 0.5 mm wide slit before entering into a decelerating lens system (close to
 ground potential) followed by a hemispherical analyzer.
 The lens system decelerates the electrons to the pass energy and focuses them at the entrance of the analyzer.
 The electrons are then detected by a two-dimensional detector after passing through a hemispherical analyzer, and the precise energy is calculated from the impact position.
 SIMION simulations \cite{dahl2000simion} show that all electrons transmitted through
 the slit will enter the analyzer (see the bottom panel in Fig.~\ref{fig:EMS_SIMION}), hence the spectrometer transmission is determined solely by the width of the entrance slit and is independent of the electron kinetic energy. This spectrometer was operated in two different modes: A high-resolution mode and a low-resolution mode.

 In the high-resolution mode the pass energy was set to 200 eV.
 The sample high-voltage was kept constant and the analyzer voltage was varied up to 1 kV.
 Stability of the sample high-voltage was checked using a precision voltage divider and a 7-digit volt meter, and found to be better than 0.2 V.
 However the absolute accuracy of the high-voltage measurement is not expected to be
 better than 5 V. The energy resolution was found to be 4.8 eV in this mode but the range of energies that can be measured was limited to $\approx$ 930 eV due to constraints on the voltage that can be applied to the analyzer.

 In the low-resolution mode the pass energy was set to 1000 eV, the analyzer voltage was kept constant and the sample high-voltage was controlled by a computer using a 16-bit DAC. Measurement of the obtained voltage in this mode showed deviations up to 8 V from the nominal voltage when the high voltage was varied between 5 kV and 35 kV, but when the voltage is varied over a smaller range (1-2 kV) the deviation was fairy constant ($\approx$ 1 V) over this range. The energy resolution was found to be 6.6 eV in this mode, and the energy range that can be measured is not restricted so that the data acquisition rate is about 5 times higher.

\begin{figure*}
 	\hspace*{-30pt}
 	\subfigure{\includegraphics[width=1.2\columnwidth,height=8.2cm]{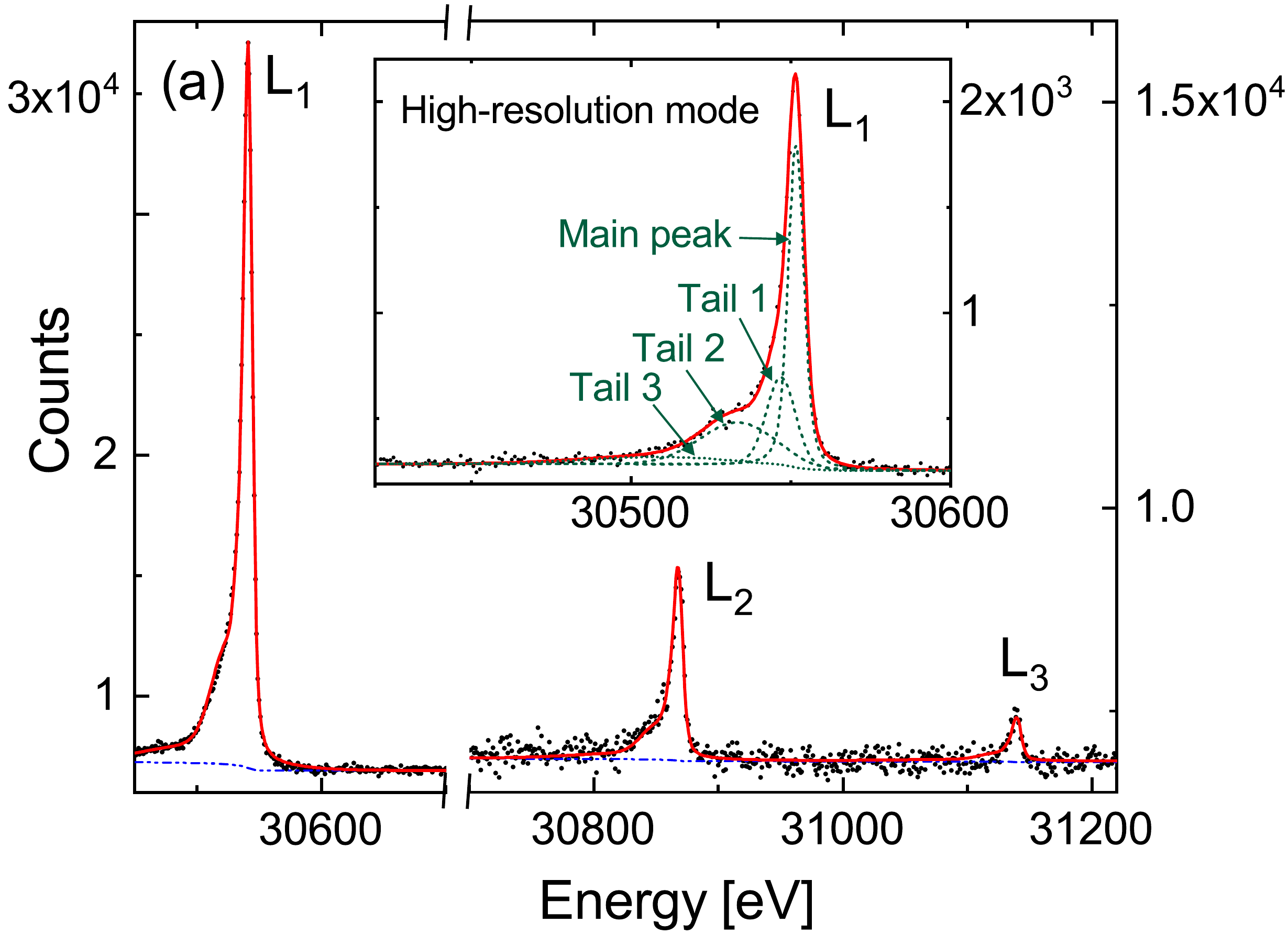}}
 	\hspace*{10pt}
 	\subfigure{\includegraphics[width=0.9\columnwidth,height=8.05cm]{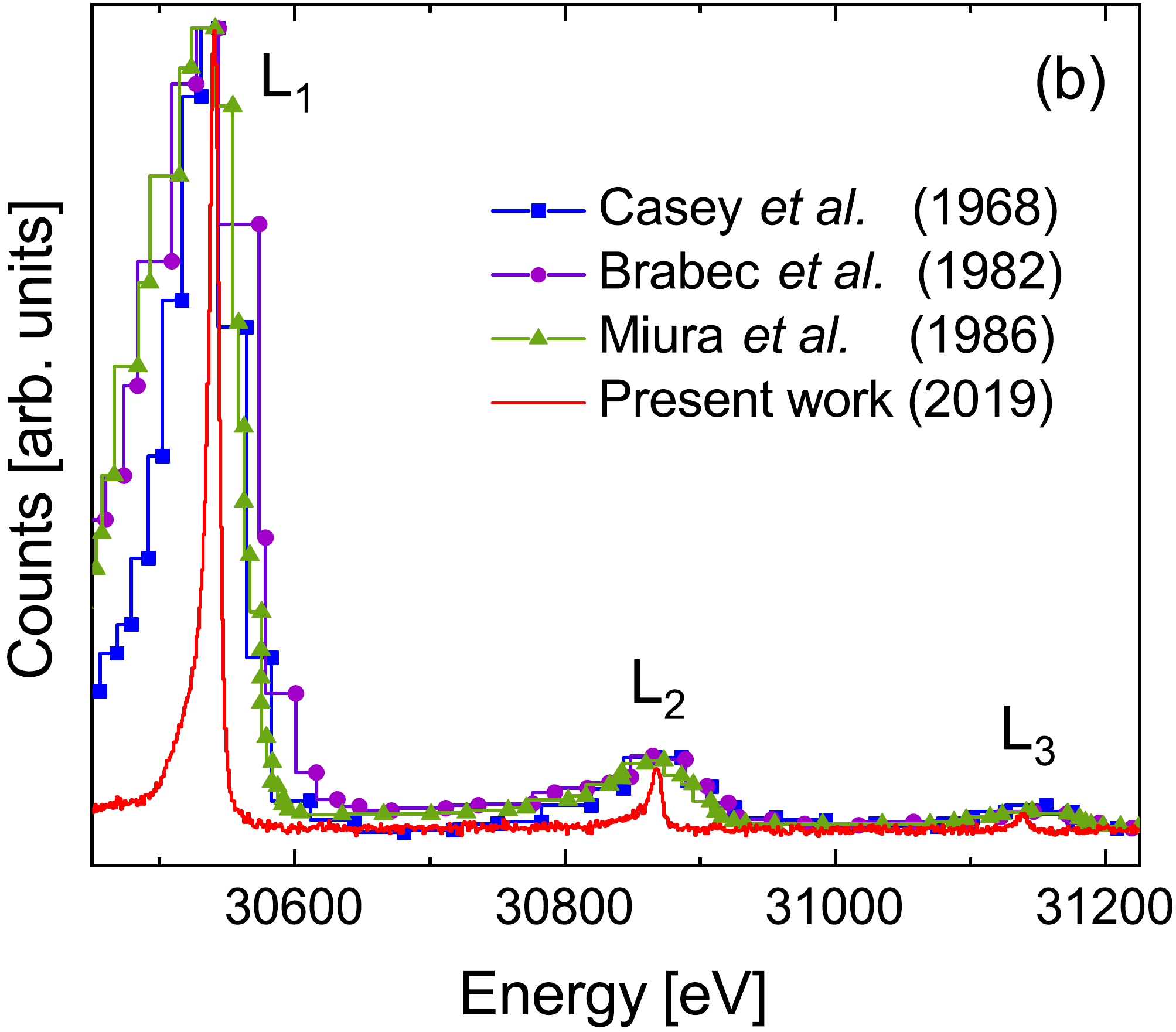}}
 	\caption{(a) The $L_1$, $L_2$ and $L_3$ conversion lines spectrum taken at 6.6 eV
 		instrumental resolution, with natural widths of 2.2 eV, 2.84 eV and 2.62 eV, respectively,
 		as adopted from Campbell \textit{et al.} \cite{campbell2001widths}.
 		The fitted curve and the background are shown in red and blue-dashed curves, respectively.
 		Note that the $L_2$ and $L_3$ lines are on a different scale to the $L_1$ line.
 		The reduced $\chi^2$ of the this fit is 1.9.
 		The $L_1$ peak measured at high resolution of 4.8 eV is shown in the insert to illustrate the fitting procedure.
 		(b) Comparison of the $L_1$, $L_2$ and $L_3$ conversion electrons of present
 		measurements (low-resolution mode) with those taken using magnetic spectrometers by Miura \textit{et al.}
 		\cite{miura1986relation}, Brabec \textit{et al.} \cite{brabec1982nuclear}, and
 		Casey \textit{et al.} \cite{casey1969thel}.
 		All previous data were digitized and scaled to match the $L_1$ peak height of the
 		present data.}
 	\label{fig:l1l32018}
 \end{figure*}

\begin{figure*}
	\centering
	\includegraphics[width=2\columnwidth,height=9cm]{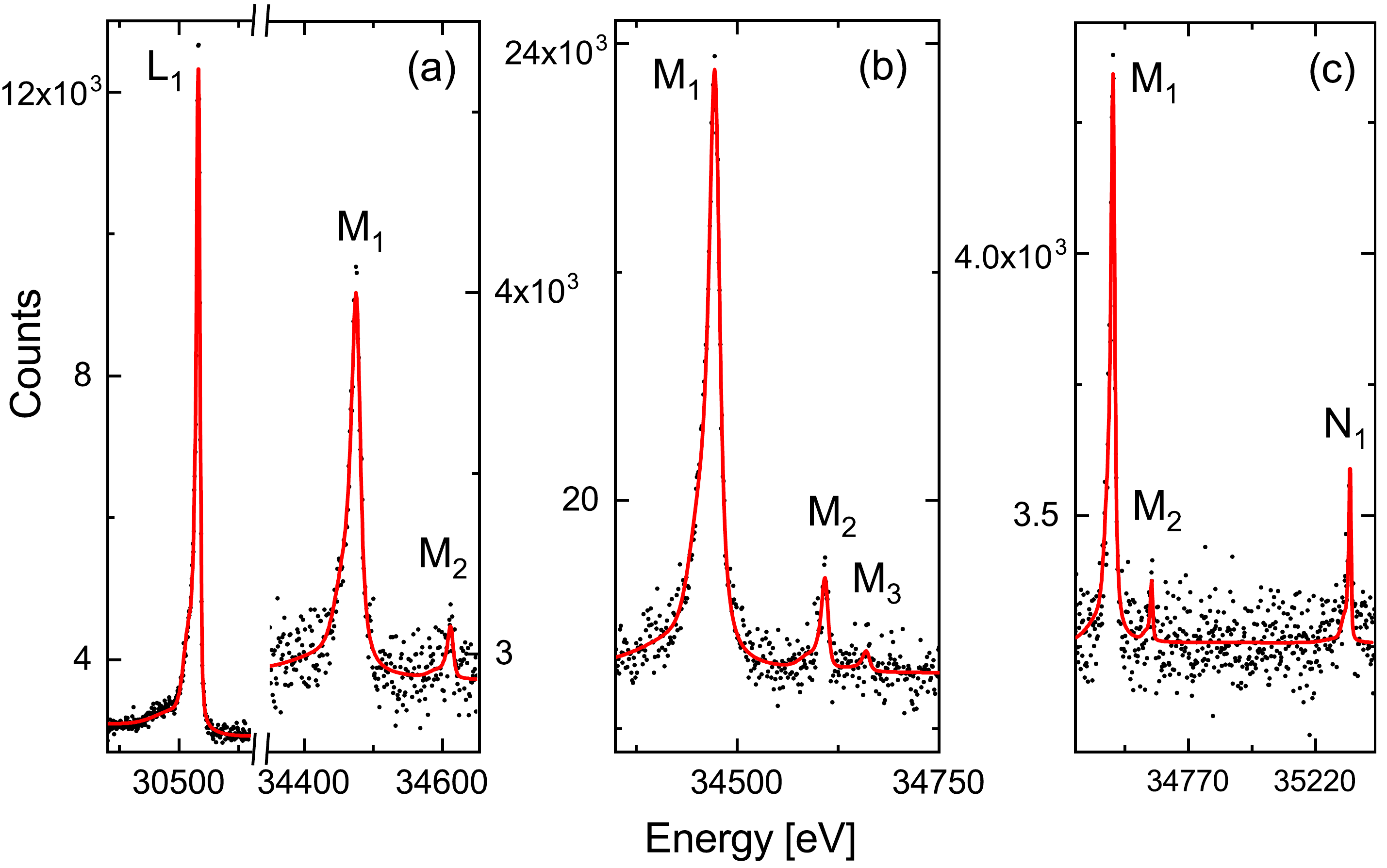}
	\vspace*{-9pt}
	\caption{ Spectra from three separate measurements in low-resolution mode:
   (a) $L_1$ conversion line measured together with $M_1$ and $M_2$.
    A 3 keV region between the peaks is omitted.
   (b) $M_1$, $M_2$ and $M_3$ conversion lines.
   (c) $M_1$, $M_2$ and $N_1$ conversion lines.
  The adopted natural widths of 10.2 eV, 3.2 eV, 3.9 eV and 2.4 eV for the $M_1$, $M_2$, $M_3$ and $N_1$ lines
  were taken from Campbell \textit{et al.} \cite{campbell2001widths}.}
	\label{fig:lmncecombinedspectra}
\end{figure*}

 Five measurements on the conversion electrons were carried out as presented
 in Fig.~\ref{fig:l1l32018} and Fig.~\ref{fig:lmncecombinedspectra}.
 Fig.~\ref{fig:l1l32018} shows the spectrum of the $L$ conversion electrons measured in two
 different modes: The $L_1$, $L_2$ and $L_3$ conversion electrons measured together in the low-resolution mode, and the $L_1$ conversion electron measured in the high-resolution mode.
 Fig.~\ref{fig:lmncecombinedspectra} shows the spectrum of the $L_1$, $M_{1,2,3}$ and $N_1$
 conversion electrons measured in the low-resolution mode and in 3 groups:
 (a) $L_1$, $M_1$ and $M_2$,
 (b) $M_1$, $M_2$ and $M_3$
 and
 (c) $M_1$ and $N_1$.

\section{Spectrum evaluation}
 All measured conversion peaks are observed to be asymmetric with a longer tail on the low-energy side. The tails generally can be attributed to energy loss of (inelastically scattered) electrons, due to intrinsic and extrinsic effects. Intrinsic effects involve a sudden change in the atomic potential due to formation of a core hole causing an outer-shell electron to be excited into another bound state or into the continuum, i.e the shake processes \cite{PhysRevA.8.2887,srivastava2009extrinsic}. Extrinsic effects involve the transportation of electrons through the solid from the emitting atom to the surface, causing inelastic scattering leading to creation of surface plasmons. Given the sample was a monolayer source, the contribution of bulk plasmons is expected to be small, and the probability of surface plasmon creation at these high energies (30 to 35 keV) is of order of 3\% \cite{dahl2000simion}. Thus it is expected that the observed tails are mostly due to the shake processes, that is, the intrinsic effects.

 The fitting strategy is as follows: Individual conversion electron peaks were fitted by convoluting a Lorentzian function with the sum of four Gaussian functions. The Lorentzian component is used to describe the lifetime broadening effects of a core shell, and the Gaussian components describe the instrumental broadening effects. Additional broadening was included in each tail component, attributed to the shake processes, by introducing a free-fitting parameter to indicate the intrinsic width of a tail in its Gaussian profile. The Gaussian widths of the tails are therefore a combination of both intrinsic tail widths, $t_w$, and instrumental resolution, $w_G$, with a magnitude of $\sqrt{t_w^2 + w_G^2}$. The Lorentzian (natural) widths of conversion lines were adopted from the latest compilation of recommended natural widths data by Campbell \textit{et al.} \cite{campbell2001widths}. In addition, a very small Shirley-type background \cite{shirley1972high} was implemented in the fit to account for the small-increment in the observed background under the peaks. A fit showing the components of a fitted line is illustrated in Fig.~\ref{fig:l1l32018}(a).

 To determine the peak area of different conversion electron lines with as few free parameters as possible, it was assumed that all conversion electrons have the same tail distributions.
 This assumption was based on the following reasons:
 \emph{(i)} The shake probabilities of the conversion electrons measured here were calculated and found to be similar \cite{1885-118287}, using the methodology proposed by Krause and Carlson \cite{krause1967vacancy}.
 \emph{(ii)} There is no reliable theory at present to describe the energy distribution of the shake electrons.
 In order to study the line shapes of the conversion electrons, the most intense ($L_1$)
 conversion line was measured in the high-resolution mode.
 From this measurement, a set of tail parameters was obtained, as summarized in
 Table~\ref{tab:tailparameters}.
 This set of tail parameters was then employed for all other conversion lines,
 by adjusting, for a given conversion peak, the Gaussian and Lorentzian widths to account
 for the different spectrometer resolutions and the lifetimes of the core shells.

\begin{table}[]
	\begin{ruledtabular}
	\caption{Adopted tail parameters in the fitting of all conversion lines}
	\begin{tabular}{lccc}
		Parameter                         & Tail\#1 & Tail\#2 & Tail\#3 \\ \hline
		Shift to the main peak (eV)       & 5       & 18      & 42      \\
		Intensity relative to main peak   & 0.4     & 0.5     & 0.2     \\
		Intrinsic width (eV)              & 8       & 24      & 55      \\
	\end{tabular}
    \end{ruledtabular}
\label{tab:tailparameters}
\end{table}

 The energies of the conversion electrons can be calculated by adopting the electron binding energies from the literature \cite{kibedi2008evaluation}.
 In an actual experiment, there are several factors that can result in deviations of the measured energies from the predicted values. These factors include whether the energies were measured relative to the Fermi level, the effect of chemical shifts, which depend on the chemical environment of the radioactive source, and differences due to the use of different
 voltage supplies in high and low-energy resolution modes.
 Indeed, these effects shift all core level energies in the same way, and the energy separations between the conversion peaks generally agree very well
 with the literature values. Therefore they were kept fixed during the fitting processes
 at the values from \cite{kibedi2008evaluation}.

 Various fits were made to the conversion electron spectra by taking natural widths from
 Campbell \textit{et al.} \cite{campbell2001widths}, Krause and Oliver \cite{krause1979natural},
 Fuggle and Alvarado \textit{et al.} \cite{fuggle1980core}, or
 from the EADL database \cite{perkins1991tables}.
 The effect of employing different fitting approaches on the measured quantities was assessed and included in the quoted uncertainties.

\section{Results}
 The results from the current conversion electron measurements are summarized in Table \ref{tab:table1},
 which also lists the literature and theoretical values.
 Note that the calculated ICCs are given for two different sets of the nuclear parameters (the $\lambda$ penetration parameter and the $\delta(E2/M1)$ mixing ratio). One set is from the previous work of Brabec et al. \cite{brabec1982nuclear}, and one from the present analysis (see Discussion below).

 The $L_1$:$L_2$:$L_3$ intensity ratios from the present measurement are $1:0.085(3):0.019(3)$.
 These are consistent with the ratios reported by Geiger \textit{et al.} \cite{GEIGER1965352} and Coursol \textit{et al.} \cite{coursol1980etude}.
 The obtained $L_1$:$L_2$ ratio appears to be larger, but consistent with, the ratios reported by Brabec \textit{et al.} \cite{brabec1982nuclear} and Casey \textit{et al.} \cite{casey1969thel}, where both studies used magnetic spectrometers.
 In comparison, the present measurement has a higher energy resolution, which is illustrated in
 the right panel of Fig.~\ref{fig:l1l32018}.
 All measured $L_1$:$L_3$ ratios are similar, except for the ratio obtained by Casey \textit{et al.} \cite{casey1969thel}, which we consider as an outlier.
 Note that Casey \textit{et al.} \cite{casey1969thel} quoted a 20\% uncertainty on their
 $L_1$:$L_2$ ratio but a 5\% error in their $L_1$:$L_3$ ratio which does not seem correct, as the $L_2$ line is much more intense than the $L_3$ line.
 The $L_1$:$M_1$ intensity ratio was found to be $1: 0.202(5)$, which is in agreement
 with the predicted ratio of $1: 0.198$.
 The $L_1$ to $M_1$ intensity ratio is not sensitive to the change in mixing ratio $\delta(E2/M1)$
 and penetration parameter $\lambda$ (see Table \ref{tab:table1}).
 Considering the energy difference between $L_1$ and $M_1$ conversion lines, the fact that the measured
 ratio is in agreement with the theoretical expectations indicates that the transmission of the
 electrostatic spectrometer was indeed not sensitive to the energy, which is in accord with SIMION simulations.
 The measured $L_1$:$M_2$  intensity ratio is $1:0.017(3)$. However, this ratio was not included in the analysis, as from the $L_1$:$M_1$ ratio and $M_1$:$M_2$:$M_3$ ratios a more accurate
 $L_1$:$M_2$ ratio can be deduced.
 The $M_1$:$M_2$:$M_3$  intensity ratios are $1:0.095(6):0.023(7)$.
 The $M_1$:$M_2$ ratio agrees broadly with the ratios reported by Coursol \textit{et al.} \cite{coursol1980etude} and
 Brabec \textit{et al.} \cite{brabec1982nuclear}.
 The $M_1$:$M_3$ intensity ratio is close to the ratio reported by Brabec \textit{et al.}
 \cite{brabec1982nuclear}, but is much larger than the value reported by Coursol \textit{et al.} \cite{coursol1980etude}.
 Finally, the $M_1$ conversion peak was measured together with the $N_1$ conversion peak.
 This measurement also revealed the $M_2$ conversion peak, however this peak was too weak to obtain
 accurate information on its intensity.
 Note that the $M_3$ conversion peak could not be observed in this case due to its low yield.
 The intensity ratio of $M_1$ and $N_1$ conversion lines measured here is $1:0.18(2)$, which is slightly larger than the ratio reported by Brabec \textit{et al.} \cite{brabec1982nuclear}.
 Nevertheless, both measured ratios are consistent with the predicted value of $1:0.198$,
 which is insensitive to the change in nuclear parameters, $\lambda$ and $\delta$.

\section{Discussion}
\subsection{Internal conversion coefficients and the nuclear parameters}
\begin{figure}
 	\centering
 	\includegraphics[width=\columnwidth,height=2\columnwidth]{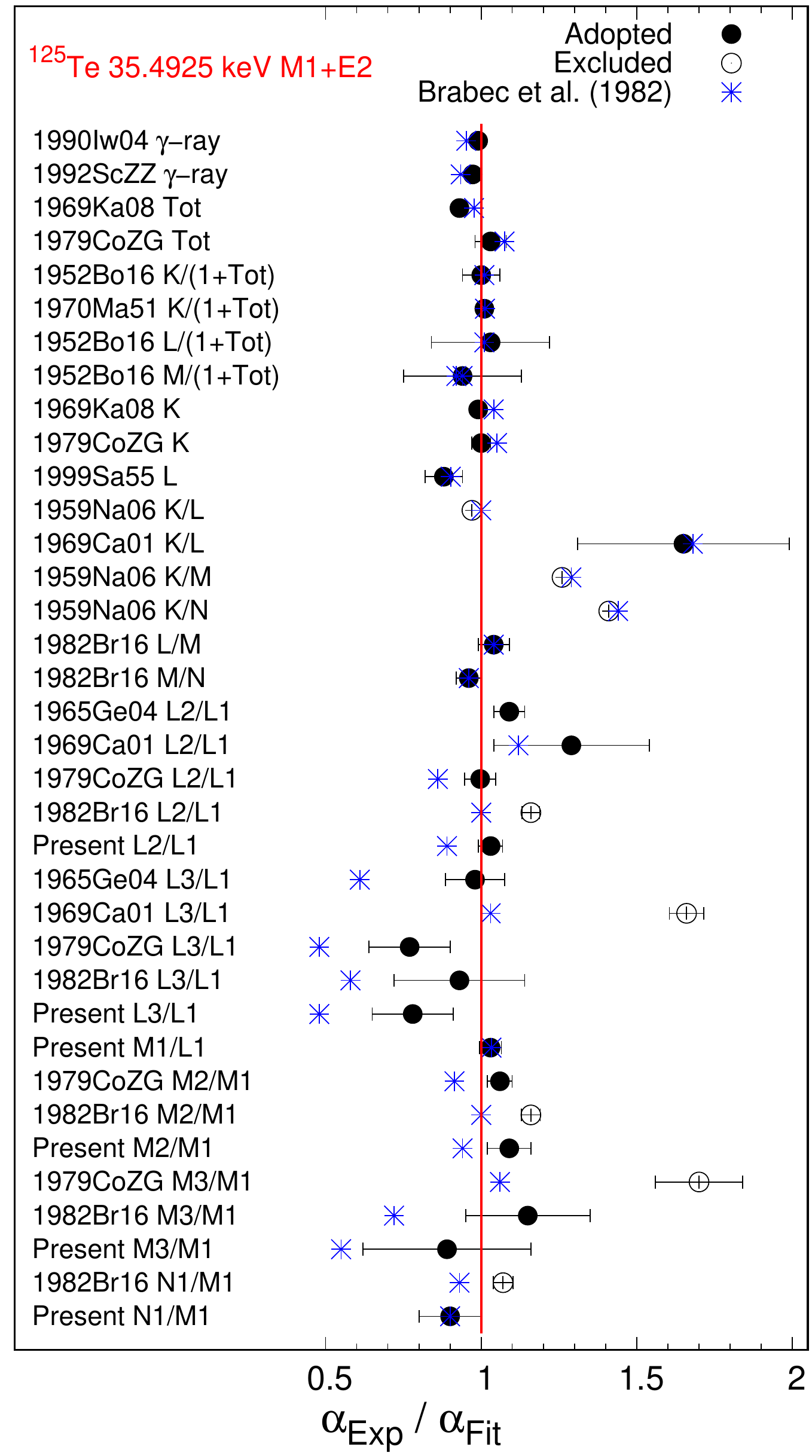}
 	\caption{Ratios of the experimental conversion coefficients to the calculated values using current evaluated nuclear parameters, $\lambda=-1.2(6)$, $\delta=0.015(2)$ (black dots), and evaluated nuclear parameters by Brabec \textit{et al.} \cite{brabec1982nuclear} in 1982 (blue stars, shown without uncertainties for clarity). White circles represent experimental values that are not included in the least-square fitting for the evaluation of current adopted nuclear parameters. The left panel shows the nuclear science reference key numbers (see references in Table \ref{tab:table1}) and the corresponding measured quantities. The data from 1959Na06 \cite{narcisi1959radioactivity} (uncertainties are not quoted in that paper) is included in this figure for comparison. Note the ratios for $\delta$ are not plotted due to their large variations and errors.}
 	\label{fig:125inuclearparameters}
 \end{figure}

 The multipolarity of the 35.5-keV transition from the decay of the $3/2^+$ excited state
 of $^{125}$Te is known to be almost pure ($>$99\%) $M1$.
 Therefore the impact of the penetration effect is considered only for the magnetic transition.
 The conversion coefficients of (sub-)shell $i$ are then related to $\lambda$ and $\delta(E2/M1)$, as
 given by Eq. (\ref{eqn:ICC_ML}) and Eq. (\ref{eq:Mixing}):
\begin{equation}
\alpha_{i,\rm{exp}} = \frac{\alpha_i(M1)(1 + b_1(i)\lambda + b_2(i)\lambda^2) + \delta^2\alpha_i(E2)}{1 + \delta^2}. \\
\label{eq:lambda_delta}
\end{equation}
\noindent Note the theoretical penetration coefficients $b_1(i)$ and $b_2(i)$ are constants for a given atomic shell.
 The conversion coefficients for pure magnetic dipole and electric quadrupole transitions of
 (sub-)shell $i$ ($\alpha_i(M1)$ and $\alpha_i(E2)$, respectively) are adopted from the BrIcc code \cite{kibedi2008evaluation}, and $b_1(i)$ and $b_2(i)$ were calculated using Dirac Hartree-Fock-Slater wavefunctions \cite{LIBERMAN1971107} in a modified version of the code CATAR \cite{pauli1975computer}.
 The adopted values of $\alpha_0(M1)$, $\alpha_0(E2)$, $b_1(i)$ and $b_2(i)$ are shown in Table~\ref{tab:table3}.
 A similar expression can also be obtained for the ratios of $i$ and $j$ sub-shell conversion coefficients:
\begin{equation}
(\frac{\alpha_i}{\alpha_j})_{\rm exp} = \frac{\alpha_i(M1)(1 + b_1(i)\lambda +
b_2(i)\lambda^2) + \delta^2\alpha_i(E2)}{\alpha_j(M1)(1 + b_1(j)\lambda +
b_2(j)\lambda^2) + \delta^2\alpha_j(E2)} \, . \\
\label{eq:ICC_ratio}
\end{equation}
\begin{table}
	\small
	\caption{\label{tab:table3}
		Theoretical internal conversion coefficients for pure $M1$ and $E2$ transitions
		(calculated using $BrIcc$ \cite{kibedi2008evaluation}) and the  $b_1$ and $b_2$ penetration coefficients
		(calculated using a modified version of the code CATAR \cite{pauli1975computer}) for the 35.5-keV transition in $^{125}$Te. }
	\begin{ruledtabular}
		\begin{tabular}{ccccc}
			Orbital shell & $\alpha_{0}(M1)$ & $\alpha_{0}(E2)$ & $b_1$ & $b_2$ \\ \hline \vspace{-2pt}
			$K$   & 1.17E+1 & 1.31E+1 & -1.31E-2 & 4.32E-5 \\
			$L_1$ & 1.41E+0 & 1.34E+0 & -1.36E-2 & 4.61E-5  \\
			$L_2$ & 1.13E-1 & 2.06E+1 & -2.04E-3 & 1.06E-6 \\
			$L_3$ & 2.83E-2 & 2.93E+1 & -1.37E-5 & 8.49E-10  \\
			$M_1$ & 2.80E-1 & 2.75E-1 & -1.36E-2 & 4.65E-5  \\
			$M_2$ & 2.37E-2 & 4.28E+0 & -2.14E-3 & 1.16E-6  \\
			$M_3$ & 5.91E-3 & 6.18E+0 & -1.46E-5 & 9.51E-10 \\
			$M_4$ & 2.65E-4 & 5.41E-2 & -3.13E-6 & 8.26E-10  \\
			$M_5$ & 1.89E-4 & 6.46E-2 & \textless 1E-10         & \textless 1E-10          \\
			$N_1$ & 5.58E-2 & 5.50E-2 & -1.36E-2 & 4.66E-5 \\
			$N_2$ & 4.39E-3 & 7.89E-1 & -2.16E-3 & 1.18E-6 \\
			$N_3$ & 1.08E-3 & 1.14E+0 & -1.47E-5 & 9.72E-10  \\
			$N_4$ & 4.14E-5 & 8.36E-3 & -3.20E-6 & 8.43E-10 \\
			$N_5$ & 2.92E-5 & 9.91E-3 & \textless 1E-10 & \textless 1E-10 \\
			$O_1$ & 6.19E-3 & 6.10E-3 & -1.36E-2 & 4.66E-5 \\
			$O_2$ & 3.42E-4 & 6.15E-2 & -2.16E-3 & 1.19E-6  \\
			$O_3$ & 8.13E-5 & 8.49E-2 & -1.47E-5 & 9.74E-10 \\
		\end{tabular}
	\end{ruledtabular}
\end{table}
\noindent The optimum values of $\lambda$ and $\delta$ can be extracted from the experimental data using a least-squares fitting method. The associated $\chi^2$ value is given by
\begin{equation}
\chi^2 = \sum_{k=1}^{n}\frac{[O_k - T(R_k, \lambda, \Delta)]^2}{\sigma_k^2},
\label{eq:leastssqfit}
\end{equation}
\noindent where $O_k$, $\sigma_k$ and $T(R_k,\lambda,\Delta)$ are the experimental quantities and the corresponding uncertainties and theoretical values, of measurement $k$, respectively. The evaluation of the least-square fitting was done by using MINUIT with the MIGRAD minimizer, and the standard errors were obtained from the MINUIT processor MINOS \cite{James:1994vla}.
\begin{table}
	\small
	\caption{\label{tab:table1} Mixing ratios, penetration parameters and internal conversion coefficients for the 35.5-keV transition in $^{\rm 125}$Te. The experimental values are compared with the previous and present evaluated sets of nuclear parameters.}
	\begin{ruledtabular}
		\begin{tabular}{ccccc}
			\multirow{3}{*}{Quantity}&
			\multirow{3}{*}{Exp.} &
			\multirow{3}{*}{Ref.} &
			\multicolumn{2}{c}{Calculated} \\  \cline{4-5} &
			&  & $\lambda$=+2.4\footnote{Previously evaluated values by Brabec \textit{et al.} \cite{brabec1982nuclear}.} & $\lambda$=-1.2\footnote{Evaluated values from the present analysis.} \\
			& & & $\abs{\delta}$=0.029 & $\abs{\delta}$=0.015  \\ \hline
			$100/(1+\alpha_{Tot})$ & 6.68(14)  & 1990Iw04\cite{iwahara1990determination} & 7.01  & 6.73 \\
			& 6.55(13) &  1992ScZZ\cite{schotzig1992precision} && \\
			$\alpha_{Tot}$ &
			13.65(28) & 1969Ka08\cite{KARTTUNEN1969343} &  13.26 & 13.86 \\
			& 14.25(64) & 1979CoZG\cite{coursol1980etude}&& \\
			$\alpha_K/(1+\alpha_{Tot})$ & 0.80(5)  & 1952Bo16\cite{PhysRev.85.858} & 0.795 & 0.800 \\
			& 0.804(10) & 1970Ma51\cite{marelius1970conversion} & & \\
			$\alpha_L/(1+\alpha_{Tot})$& 0.11(2) &  1952Bo16\cite{PhysRev.85.858} & 0.109  & 0.107 \\
			$\alpha_M/(1+\alpha_{Tot})$& 0.020(4) &  1952Bo16\cite{PhysRev.85.858} & 0.022 & 0.021 \\
			$K$ & 12.01(18)  & 1969Ka08\cite{KARTTUNEN1969343} & 11.33 & 11.88 \\
			& 11.90(31) & 1979CoZG\cite{coursol1980etude} && \\
			$L$& 1.4(1)  &  1999Sa55\cite{PhysRevC.60.064309} & 1.55  & 1.59 \\
			$K/L$& 12.3(25)  &  1969Ca01\cite{casey1969thel} & 7.30  & 7.47 \\
			$L/M$& 5.21(26)  &  1982Br16\cite{brabec1982nuclear} & 5.00  & 5.00 \\
			$M/N$& 4.87(20)  &  1982Br16\cite{brabec1982nuclear} & 5.08  & 5.07 \\
			$L_2/L_1$ &  0.089(4)  &  1965Ge04\cite{GEIGER1965352} & 0.095 & 0.082 \\
			& 0.106(21) &  1969Ca01\cite{casey1969thel} & &  \\
			& 0.082(4) &  1979CoZG\cite{coursol1980etude} & & \\
			& 0.095(2) \footnote{\label{excluded}Excluded in the present least-square fitting analysis.} &  1982Br16\cite{brabec1982nuclear} & & \\
			&  0.085(3) & Present  & & \\
			$L_3/L_1$ &  0.024(2)  &  1965Ge04\cite{GEIGER1965352} & 0.039 & 0.025 \\
			&  0.041(2) \footref{excluded} &  1969Ca01\cite{casey1969thel} & &  \\
			&  0.019(3) &  1979CoZG\cite{coursol1980etude} & & \\
			&  0.023(5)  &  1982Br16\cite{brabec1982nuclear} & & \\
			&  0.019(3) & Present  & & \\				
			$M_1/L_1$   & 0.202(5)& Present  & 0.198 & 0.198  \\
			$M_2/M_1$   & 0.092(5)  &  1979CoZG\cite{coursol1980etude} & 0.101  & 0.087  \\
			& 0.101(5) \footref{excluded} & 1982Br16\cite{brabec1982nuclear} & & \\
			& 0.095(6)  & Present  & & \\
			$M_3/M_1$ & 0.044(3) \footref{excluded}  &  1979CoZG\cite{coursol1980etude} & 0.042 & 0.026 \\
			& 0.030(5) & 1982Br16\cite{brabec1982nuclear} & & \\
			& 0.023(7) & Present  & & \\
			$N_1/M_1$     &  0.214(6) \footref{excluded} & 1982Br16\cite{brabec1982nuclear} & 0.199 & 0.199   \\
			&  0.18(2) & Present  &  &    \\
			$N_1/M_1$     &  0.214(6) & 1982Br16\cite{brabec1982nuclear} & 0.199 & 0.199   \\
			$\delta$ \footnote{The quoted experimental values from the angular distribution and correlation measurements are from the compilation by Krane \cite{KRANE1977363}.}      &  +0.09(1) \footref{excluded} & 1971Ba44\cite{BARRETTE1971101}  & 0.029  & 0.015   \\
			&  +0.095(25) \footref{excluded} & 1971Ba44\cite{BARRETTE1971101}  &  &    \\
			&  +0.078(12) \footref{excluded} & 1971Ba44\cite{BARRETTE1971101}  &  &    \\
			&  +0.08(3) \footref{excluded} & 1971Wy02\cite{PhysRevC.3.2442}  &  &    \\
			&  +0.04(8)     & 1971Wy02\cite{PhysRevC.3.2442}  &  &    \\
			&  +0.12(7)     & 1971Wy02\cite{PhysRevC.3.2442}  &  &    \\
			&  -0.002(16)   & 1972Ba12\cite{badica1972gamma}  &  &    \\
			&  -0.02(13)    & 1972Ba12\cite{badica1972gamma}  &  &    \\
			&  +0.04(6)     & 1972Ba12\cite{badica1972gamma}  &  &    \\
		\end{tabular}
	\end{ruledtabular}
\end{table}
 Using Eq. (\ref{eq:lambda_delta}), Eq. (\ref{eq:ICC_ratio}) and Eq. (\ref{eq:leastssqfit}), and the current and previous experimental values from Table~\ref{tab:table1}, the best-fit values of $\lambda=-1.2(6)$ and $\abs{\delta}=0.015(2)$ were obtained, with a reduced $\chi^2=1.2$. Some literature values were excluded in the least-square approach, since they are more than two standard deviations away from the corresponding fitted values (see Table~\ref{tab:table1} and Fig.~\ref{fig:125inuclearparameters}). The ratios of the experimental values to the calculated values with this set of nuclear parameters is shown in Fig.~\ref{fig:125inuclearparameters}. The figure also compares the current evaluated values with the previously evaluated values determined by Brabec \textit{et al.} \cite{brabec1982nuclear}.
 The $\lambda=-1.2(6)$ value obtained in the present analysis indicates a smaller anomaly in the conversion coefficients than reported previously. In order to compare this value with theory, we have calculated $\lambda$ using the intermediate coupling approach in the particle-vibrational (PV) model, in which an odd neutron is coupled to the quadrupole vibrations of a spherical core, and the coupling strength is described by a dimensionless parameter $\xi$. The formulations can be found in Ref. \cite{choudhury1967intermediate,HEYDE196781,berghe1970penetration}. In the calculations, the 2$s_{1/2}$, 1$d_{3/2}$, 1$d_{5/2}$ and 0$g_{7/2}$ shell-model orbits were considered and the extra-core nucleon was coupled to one and two phonon core excitations. Radial integrals over the reduced coordinate $(r/R_0)$ were evaluated with harmonic oscillator wavefunctions with phase chosen so that the wavefunctions are positive as $r \rightarrow \infty$. This phase convention is implicit in the formulation of the PV model Hamiltonian in that the coupling parameter $\xi$ is positive for all single-particle orbits. The renormalization of the $g_s^{(b)}$ factor was chosen to be 0.6 according to the results in \cite{listengarten1976renormalization}, and it was assumed that $g_s^{(a)}=g_s^{(M1)}=g_s^{(b)}$. This assumption is reasonable for the hindered $M1$ transitions since the terms associated with the $g_s^{(a)}$ and $g_s^{(M1)}$ factors in Eq. (\ref{eq:PE_GAMMA_MATRIX}) are small relative to the term associated with $g_s^{(b)}$.
 \begin{table}[]
 	\small
 	\caption{Particle vibrational (PV) model results for the 35.5-keV $\gamma$ transition and the states involved in $^{\rm 125}$Te}
 	\begin{ruledtabular}
 		\begin{tabular}{ccc}
 			Quantity & PV model \footnote{Particle-vibrational model, present calculations.} & Exp. \footnote{The quoted $g$ factors are from \cite{PhysRevC.80.054301}, and the other values except $\lambda$ from \cite{KATAKURA2011495}. $\lambda$ is from present analysis.}\\ \hline
 			$E_x$($3/2^+$)[keV] & 36 & 35\\
 			$g(1/2^+)$ & -1.85 & -1.78 \\
 			$g(3/2^+)$ & +0.580 & +0.403(3) \\
 			$B(M1;3/2^+$$\rightarrow$$1/2^+)$[W.u.] & 0.0004 & 0.0226(4) \\
 			$B(E2;3/2^+$$\rightarrow$$1/2^+)$[\rm W.u.] & 15.6 & 11.9(24) \\
 			$U_\gamma(M1)$[$\rm\mu_N$] & -0.054 & (+)0.402(4) \footnote{\label{signlambda}The sign of this quantity was deduced semi empirically as described in the main text.} \\
 			$M_{e1}$[$\rm \mu_N$] & -0.37 & - \\
 			$M_{e2}$[$\rm \mu_N$] & -0.96 & - \\
 			$\lambda$ & +5.57 & -1.2(6) \\
 			          &  (-)0.8 \footnote{Evaluated with the experimental value of $U_\gamma(M1)$.} & \\
 		\end{tabular}
 	\end{ruledtabular}
 	\label{tab:PVresults}
 \end{table}
 The following $g$ factors: $g_s^{(b)}=0.6 \times g_s^{free}$, $g_l=0.0$, $g_R=0.4$ were used and the coupling strength parameter was taken to be $\xi=3.0$, which is reasonable and near maximal for nuclei in this mass region \cite{HEYDE196781}. The calculation results are summarized and compared in Table \ref{tab:PVresults}. The calculated $g$ factors of the 1/2$^+$ and 3/2$^+$ states are close to their experimental values, and the calculated $E2$ transition strength between these two states is also consistent with the experimental value. This agreement indicates that the PV model describes the dominant part of the 1/2$^+$ and 3/2$^+$ state wavefunctions reasonably well. As a consequence of the small $M1$ strength, the magnitude of the calculated $\lambda=+5.6$ is about five times larger than the experimental value, and the predicted sign is not in accord with our analysis. The overestimation of $\lambda$ stems from the underestimation of $U_\gamma(M1)$ (or equivalently the $B(M1)$) as indicated in Table \ref{tab:PVresults}. Thus the PV basis is clearly not sufficient to describe $U_\gamma(M1)$ accurately. The main contribution to the calculated $U_\gamma(M1)$ comes from the configuration mixing of higher states into the transition states. Also, additional currents such as meson exchange and/or velocity dependent forces, which are usually small for allowed transitions \cite{church1960nuclear}, are not being considered in the formulated $U_\gamma(M1)$ operator in Eq.~(\ref{eq:PE_GAMMA_MATRIX}). Such small currents could possibly explain the observed discrepancy. Since the penetration matrix elements $M_{e1}$ and $M_{e2}$ are not sensitive to the choice of parameters (except the $g_s$ factor), a common strategy in the literature has been to deduce $\abs{U_\gamma(M1)}$ from the experimental value of the reduced $B(M1)$ probability using the relation \cite{listengarten1978penetration}:
 \begin{equation}
 |U_\gamma(M1)|^2 =(2I_i + 1)B(M1) \ ,
 \label{eq:Ugamma_BM1}
 \end{equation}
 \noindent where $I_i$ is the spin of the initial state.
 Note that in Eq. (\ref{eq:Ugamma_BM1}), only the magnitude of $U_\gamma(M1)$ can be obtained. In this way we obtained $\abs{\lambda}=0.8$, which agrees with the magnitude of the experimental $\lambda$. In order to predict the sign of $\lambda$, we compared the positive sign of the mixing ratio $\delta(E2/M1)$ from the angular distribution and correlation results in the literature \cite{KRANE1977363} with the calculated positive sign of the $E2$ gamma matrix element. Using the relation described in Eq.~(\ref{eq:mixing_ratio}), $U_\gamma(M1)$ is then deduced to be positive. Thus, since the calculated $M_{e}$ is negative, $\lambda$ is deduced to be negative, which is in accord with our experimental results. This semi-empirical analysis clearly requires the reliability of the calculated signs of $E2$ gamma matrix element and penetration matrix elements $M_{e1}$, $M_{e2}$. Care has taken to use consistent phase conventions. Given that these matrix elements are allowed, the calculated signs of these quantities should be reasonably reliable.

 Other theoretical calculations suggest $\abs{\lambda}=1.2$ (single-particle model \cite{listengarten1976renormalization}), $\abs{\lambda}=1.4$ (finite Fermi systems theory \cite{kopytin1978conversion}) and  $\lambda= +0.7$ to $+0.8$ \cite{rao1975internal} (microscopic core-polarization theories using an effective $M1$ operator). Note that in \cite{rao1975internal}, the core-polarization theories require the sign of $\lambda$ to be positive for both odd proton and odd neutron transitions, which applies to our case in $^{125m}$Te ($\nu_{d_{3/2}}$$\rightarrow$$\nu_{s_{1/2}}$). However this is not what we observed from the optimum value of $\lambda$. If the evaluated $\lambda$ is limited to be strictly positive, the optimum nuclear parameters become $\lambda=0.0(+7)$ and $\delta=0.016(2)$, with a reduced $\chi^2=1.8$.

 It is instructive to compare the measured $\lambda$ for the $l$-forbidden $M1$+$E2$ transitions in $^{121}$Te, $^{123}$Te and $^{125}$Te, as shown in Table \ref{tab:lambda_for_Te_isotopes}. The sign of the penetration parameters for a sequence of isotopes is likely to be the same for the same type of transition, as having a change in sign would imply a significant change in the corresponding state wavefunction across the isotopes \cite{berghe1970penetration,giannatiempo1984penetration}. In Table \ref{tab:lambda_for_Te_isotopes}, all measured $\lambda$ values are small, and are consistent with the predicted values. The comparison shows that the sign of $\lambda$ from these isotopes tends to be negative; however, this does not exclude the possibility of having positive $\lambda$. Using the same semi-empirical analysis of deducing the sign of $\lambda$ in $^{\rm 125}$Te, the signs of $\lambda$ in $^{121}$Te and $^{123}$Te are deduced to be negative. Thus our results suggest the core-polarization theory does not describe the $l$-forbidden $\nu_{d_{3/2}}$$\rightarrow$$\nu_{s_{1/2}}$ transitions in the Te isotopes correctly. More precise measurements on the $^{121}$Te and $^{123}$Te isotopes could help to draw a firm conclusion on this problem.
 \begin{table*}[]
 	\begin{ruledtabular}
 	\small
 	\caption{Comparison of the penetration parameter $\lambda$ for the $M1$+$E2$ transition ($\nu_{d_{3/2}}$$\rightarrow$$\nu_{s_{1/2}}$) for the $^{121,123,125}$Te isotopes.}
 	\begin{tabular}{cccccccc}
 		\multirow{2}{*}{Nucleus} \ & \ \multirow{2}{*}{$\gamma$ energy (keV)} \ & \ \multirow{2}{*}{$\abs{\delta(E2/M1)}$}\ & \ \multirow{2}{*}{Measured $\lambda$} \   &  \multicolumn{4}{c}{Calculated $\abs{\lambda}$} \\ \cline{5-8}
 		& & & & \ Ref \cite{listengarten1976renormalization} \footnote{\label{1stfoot} Evaluated with effective $g$ factor $g_s^{\rm eff}$ = 0.6 $\times$ $g_s^{free}$.} \   & \ Ref \cite{kopytin1978conversion} \footref{1stfoot} & \ Ref \cite{rao1975internal} \ & \ PV model \footnote{ Particle-vibration model. Present calculations.} \ \\ \hline
 		$^{121}$Te   	& 212.2    & 0.226(8) \cite{OHYA20101619} & -0.7(17) \cite{edvardson1971total} & 1.0  & 1.1  & 0.6 to 0.8 & 0.6 \\
 		& & & $-$3 to +4 \cite{NSR1973SA46} & & & &  \\
 		$^{123}$Te		& 159.0   &	   0.062(6) \cite{OHYA2004547}   &   $-$2(2) \cite{tornkvist1969k}     & 1.2        & 1.3 & 0.6 to 0.8 & 0.8  \\
 		$^{125}$Te		& 35.5     &	0.015(2)   &	$-$1.2(6)   & 1.2  & 1.4  &  0.7 to 0.8  & 0.8
 	\end{tabular}
    \label{tab:lambda_for_Te_isotopes}
    \end{ruledtabular}
 \end{table*}

 The evaluated mixing ratio of $\abs{\delta}=0.015(2)$ is in accord with $\abs{\delta}=0.019$, predicted by \cite{badica1972gamma}, using the Kisslinger-S{\"{o}}rensen model.
 The sign of this quantity cannot be deduced from ICC measurements because of the quadratic
 dependence on $\delta$ in Eq. (\ref{eq:Mixing}).

\subsection{Shake processes}
 The observed tails on the conversion lines are expected to be due to the shake processes.
 Our high-resolution measurements provide an opportunity to study the shake electrons emitted from outer shells.
 Assuming that the tails correspond to shake electrons,
 using Table~\ref{tab:tailparameters}, the shake probability for Te can be estimated
 from the fitted tail intensities: 100\%$\times$(0.4+0.5+0.2)/(1+0.4+0.5+0.2) $\approx$50\%
 of the total peak area.
 This value is more than two times larger than the predicted value of $\approx$20\%,
 obtained from a calculation based on the single-configuration framework \cite{1885-118287}.
 It should be noted that this model does not take into account the electron-electron correlations.
 There are at least two possible explanations for the discrepancy between expected and observed shake intensity:
 \emph{(i)} the large overlap between the tail and the main peak may have overestimated the tail intensity,
 and/or
 \emph{(ii)} the shake probability calculations based on the single-configuration framework may underestimate the effect.
 Calculations that exclude the electron-electron correlations are mainly valid
 for closed-shell atoms.
 Lowe \textit{et al.} \cite{lowe2011ab} demonstrated that the inclusion of
 electron-electron correlations in the calculations of shake probability describe better the transition
 metals (which all have an open $3d$ shell).
 Their calculated probabilities are up to seven-times larger than the ones based on the single-configuration
 framework.
 Tellurium has six valence electrons and atomic configuration of $5s_{1/2}^25p_{1/2}^25p_{3/2}^2$,
 i.e it has an open $5p$ shell, which might likewise signal the importance of correlations and help explain the observed high shake probability.

 The energy distribution of the shake electrons associated with a transition of energy $E_\gamma$ can be described using the following equation \cite{krause1967vacancy}:
 \begin{equation}
 E_{\rm shake} + E_{\rm core} = E_\gamma - E_i - E_j^*,
 \label{eq:shake}
 \end{equation}
 \noindent where $E_\gamma = 35.5$ keV, $E_{\rm shake}$ and $E_{\rm core}$ are the kinetic energies of the shake and core electrons, respectively. $E_i$ is the binding energy of (sub-)shell $i$ (where the core electron is emitted), and $E_j^*$ is the binding energy of (sub-)shell $j$ (where the shake electron is emitted) with a vacancy present in (sub-)shell $i$. Note that $E_j^*$ is approximately the binding energy of (sub-)shell $j$, and in this case the core electron is the conversion electron. The emitted conversion electron and shake electron are indistinguishable, hence the result is a continuous energy distribution from zero energy up to $E_\gamma - E_i - E_j^*$. Since the shake electrons are mostly from the $N_4$, $N_5$, $O_1$, $O_2$ and $O_3$ subshells \cite{1885-118287}, which have binding energies of 40.8 eV, 39.2 eV, 11.6 eV, 2.6 eV and 2.0 eV, respectively \cite{kibedi2008evaluation}, the tail shift parameters in Table~\ref{tab:tailparameters} are of the correct magnitude to correspond to these binding energies. Differences between the fit energies and the outer-shell binding energies may be influenced by the use of a symmetrical (Gaussian) peak shape in the fit, whereas the actual energy distribution of the shake electrons is expected to be asymmetric \cite{mukoyama2005electron}. Further investigations are needed to make a more quantitative evaluation of the shake processes.

\section{Conclusion}
 High resolution electron spectroscopy following the electron-capture decay of $^{125}$I has been reported. By combining the present and literature values of conversion electron intensity ratios, we have evaluated new values of the penetration parameter, $\lambda=-1.2(6)$, and the $E2/M1$ mixing ratio, $\abs{\delta}=0.015(2)$, for the 35.5-keV ($\nu_{d_{3/2}} \rightarrow \nu_{s_{1/2}}$) transition in $^{125}$Te. The magnitude of $\lambda$ is consistent with our calculated $\abs{\lambda}=0.8$ using the particle vibrational (PV) model with the experimental $U_\gamma(M1)$, and is also consistent with other the theoretical values using alternative nuclear models \cite{listengarten1976renormalization,kopytin1978conversion,rao1975internal}. The negative sign of $\lambda$ is not consistent with the theoretical prediction of \cite{rao1975internal} which adopts a core-polarization approach, whereas it agrees with our semi-empirical analysis based on the sign of the mixing ratio \cite{KATAKURA2011495} and the calculated sign of the $E2$ gamma matrix element. Nonetheless, since $\lambda$ is small, the penetration effect on the internal conversion coefficients is less than 4$\%$ for this case. The obtained $\abs{\delta}$ is in agreement with the theoretical prediction in \cite{badica1972gamma}.
 \raggedbottom

 The electron shake processes arising from the emission of the conversion electrons has also been investigated. It was found that the measured shake probability for the $L$, $M$ and $N$ conversion electrons is about 50\%, which is 2.5 times larger than the predicted value of 20\%, based on single-configuration calculations \cite{1885-118287}. Our results may indicate the importance of the inclusion of electron-electron correlations in the shake-probability calculations for an open shell atom like Te.

 As noted in the introduction, a primary aim of the present measurements was to determine the Auger electron yields for the medical isotope $^{125}$I by simultaneously measuring the Auger electron yields relative to the conversion electron yields. The present paper has focused on the nuclear parameters $\lambda$ and $\delta$, which affect the conversion electron yields of the relevant $l$-forbidden 35.5 keV nuclear $M1$ transition in $^{125}$Te. The Auger electron yields have been published elsewhere \cite{alotiby2018measurement,ALOTIBY2019} and compared with results of computational models \cite{BrIccEmismodel,PhysRevA.21.442}. These recently published Auger yields were based on conversion electron yields evaluated with the nuclear parameters determined here.

 Looking to future experimental evaluations of Auger yields from radioisotopes by this method, it appears that penetration effects are generally small for spherical nuclei in this mass region \cite{listengarten1976renormalization,rao1975internal}. However, much larger penetration factors have been reported in some cases \cite{GERHOLM1965441}. Because the measurement of penetration factors is so difficult, it is important to have a reliable estimate on whether they are significant or not. The present theoretical analysis suggests that for $l$-forbidden $M1$ transitions, a useful estimate of the magnitude can be obtained by combining a theoretical model for the allowed electron penetration matrix elements with the experimental forbidden $M1$ $\gamma$-radiation matrix element.

\section{Acknowledgement}
This research was made possible by an Australian Research Council Discovery Grant DP140103317. J.T.H. Dowie acknowledges support of the Australian Government Research Training Program.

\newpage
\bibliography{125I_references}

\end{document}